\documentclass[format=sigconf,nonacm]{acmart}

\usepackage[utf8]{inputenc}
\usepackage[T1]{fontenc}
\usepackage{microtype}
\usepackage{lmodern}
\usepackage{courier}
\usepackage{graphicx}
\usepackage{xcolor}
\usepackage{float}
\usepackage{subcaption}
\usepackage[hang]{footmisc}
\usepackage{caption}
\usepackage{enumitem}
\usepackage{listings}
\usepackage{pgfplots}
\pgfplotsset{compat=1.18}
\usepackage{pgfplotstable}
\usepackage{ifthen}
\definecolor{linkcolor}{RGB}{31,31,222}
\definecolor{keywordcolor}{rgb}{0,0,0}
\definecolor{programcolor}{rgb}{0,0,0}
\definecolor{variablecolor}{rgb}{0,0,0}
\definecolor{bgcolor}{named}{white}
\usepackage[nameinlink]{cleveref}

\floatstyle{ruled}
\newfloat{lstfloat}{hbt}{lop}
\floatname{lstfloat}{Program}
\crefname{lstfloat}{Program}{Programs}
\crefname{lstlisting}{Listing}{Listings}
\crefname{figure}{figure}{figures}

\lstnewenvironment{python}[1][2]{%
  \lstset{language=Python, gobble=#1, numbers=left, xleftmargin=2.5em}%
}{}

\def\WebTP{WebTigerPython}

\title[\WebTP{} -- A Low-Floor High-Ceiling Python IDE for the Browser]{\WebTP{}}
\subtitle{A Low-Floor High-Ceiling Python IDE for the Browser}

\author{Clemens Bachmann}
\authornote{Both authors contributed equally to the paper.}
\orcid{0009-0003-1708-1708}
\email{clemens.bachmann@inf.ethz.ch}
\affiliation{%
  \department{Department of Computer Science}
  \institution{ETH~Zurich}
  \city{Zurich}
  \country{Switzerland}
}
\author{Alexandra Maximova}
\authornotemark[1]
\orcid{0000-0003-2598-0810}
\email{alexandra.maximova@inf.ethz.ch}
\affiliation{%
  \department{Department of Computer Science}
  \institution{ETH~Zurich}
  \city{Zurich}
  \country{Switzerland}
}
\author{Tobias Kohn}
\orcid{0000-0002-9251-8944}
\email{tobias.kohn@kit.edu}
\affiliation{%
  \department{Department of Computer Science}
  \institution{Karlsruhe Institute of Technology}
  \city{Karlsruhe}
  \country{Germany}
}
\author{Dennis Komm}
\email{dennis.komm@inf.ethz.ch}
\orcid{0000-0002-9024-1558}
\affiliation{%
  \department{Department of Computer Science}
  \institution{ETH~Zurich}
  \city{Zurich}
  \country{Switzerland}
}



\begin{document}

\begin{abstract}
With the large diversity of platforms and devices used by students, web applications increasingly suggest themselves as the solution of choice.  Developing adequate educational programming environments in the browser, however, remains a challenge and often involves trade-offs between desired functionalities and navigating the limitations of web applications, in particular the blocking single-threaded execution model.  We introduce a fully browser-based Python programming environment that explores the possibilities and demonstrates that a web application can indeed support a rich and mature set of features, ranging from Turtle graphics over educational robotics to data processing.
\end{abstract}

\maketitle

\section{Introduction}

There is a wealth of programming environments that offer a stimulating entry for novice programmers.  Especially in a K--12 context, however, software that has to be installed and run locally on the students' computers brings some non-trivial challenges, in particular since more and more schools implement a ``bring your own device'' (BYOD) strategy or use web-oriented devices. As a result, an increasing number of educational tools, including programming IDEs, are browser-based, so that no software is needed on the users' end, except for a web browser that is connected to the internet.

We introduce
\WebTP{},
a browser-based Python IDE, which is built on Pyodide~\cite{36}.  Pyodide is a port of CPython to WebAssembly and runs all Python code directly inside the browser.  \WebTP{} does this in a web worker, as to avoid the common freezing of the UI~\cite{10}.  Using PixiJS, we implemented a graphics engine that includes Turtle graphics~\cite{12} as well as support for Matplotlib.  Moreover, \WebTP{} supports executing programs on single-board computers such as the micro:bit, Calliope mini, and the corresponding Maqueen and Calli:bot robots directly using WebUSB.

\WebTP{} demonstrates the feasibility of providing fully fledged educational programming environments as a web application that may run on virtually any device.  In particular, our main contributions are:

\begin{itemize}[leftmargin=1em]
\item\textit{Fully Animated Graphics.}
We solved the problem of integrating smooth animation in the form of Turtle graphics with Pyodide, addressing the long-standing issue of dealing with the strictly single-threaded and blocking nature of JavaScript in the browser.

\item\textit{Multimedia.}
We translated the full stack of our desktop-based Python libraries to work in the browser---including graphics, animation, audio, and robotics---, allowing for existing textbooks to be used with new (web-oriented) devices such as tablet computers.

\item\textit{High Ceiling.}
We integrated support for more advanced Python libraries such as NumPy and Matplotlib, which raises the ceiling of the IDE and demonstrates that a single implementation can serve a wide range of applications and educational needs.
\end{itemize}

\section{Related Work}\label{sec:related}

Programming environments with novice K--12 programmers as target audience, such as Greenfoot for Java~\cite{29}, or block-based approaches like Alice~\cite{14}, Scratch~\cite{38}, or Snap~\cite{33}, have a long history in the educational landscape~\cite{37}.

In this paper we focus on Python as the teaching programming language.  There is a large number of educational Python IDEs that need to be installed locally, e.g., Mu~\cite{44}, TigerJython~\cite{28}, or Thonny~\cite{7}, the latter two of which come also with support for educational robotics.  Professional tools like PyCharm~\cite{41} and Jupyter Notebooks are also widely used in educational settings~\cite{26}.
However, with the proliferation of tablet computers in schools and the diversity of devices and platforms caused by BYOD policies, installing such software tools in school settings becomes increasingly hard or even infeasible.

Web applications promise a solution to this dilemma by leveraging the ubiquitous browsers as a unified platform.
Early web applications acted primarily as an interface to software run on a server, and there are still web-based programming IDEs that follow this approach.
IDEs that execute the program code on a server, such as Replit.com~\cite{5} and the Online Python Tutor~\cite{21}, may choose to run the entire program to its completion, collect the output, and then display it in the browser.  In the case of the Python Tutor, this comes with the added benefit of being able to step forward and backward in time through the program execution.  Actual user interaction, however, is impossible in such a setting, although this might be partially solved by systems such as remote desktops, as long as latency is not an issue.  Moreover, server-based solutions are expensive~\cite{25}, come with security concerns, and require a stable internet connection.

In recent years, JavaScript as a platform has matured to a degree where web applications are increasingly run directly inside the browser.  WebAssembly~\cite{22} has further strengthened this trend towards \emph{client-side} execution through better support for applications written in C and similar languages.  Unfortunately, the translation process from desktop to web application is not quite as straightforward due to the strictly single-threaded and blocking model of JavaScript execution, which causes naïvely translated applications to block or freeze the user interface (UI)~\cite{10}.  Running an application isolated in a web worker addresses this problem, but prohibits the application from any unmediated user interaction by restricting access to the UI (including the DOM).

A first generation of web-based Python IDEs relied on reimplementations of Python in JavaScript, such as Brython~\cite{3} or Skulpt~\cite{20}, where Python code might be translated to JavaScript for subsequent execution.  This approach has been used by  CodeSkulptor~\cite{39}, the CMU Code Academy~\cite{42}, Pythy~\cite{5}, Trinket~\cite{6}, Strype~\cite{47}, and WebTigerJython~\cite{45}.  Besides performance issues, re-implementations face the issue of keeping up with the development of the Python language itself and lack support for packages such as, e.g., NumPy.

Heralded by projects such as PyPy.js~\cite{27}, the second generation of web applications relies on cross compilation of existing code bases in system languages like C to WebAssembly.  This approach allows to use the actual CPython interpreter compiled to WebAssembly as done by, e.g., Pyodide~\cite{36}.  A number of programming environments build on top of this approach, e.g., Basthon~\cite{2}, FutureCoder~\cite{23}, JupyterLite~\cite{4}, Papyros~\cite{46}, PyodideU~\cite{25}, as well as \WebTP{}, which we present in this paper.

In its current form, WebAssembly does not directly support asynchronous programming.  A Python program run with Pyodide may thus block the user interface and has highly limited user interaction capabilities.  PyodideU~\cite{25} offers two possible solutions to this issue.  When run inside the browser's main thread, the Python interpreter itself is regularly interrupted by inserting exceptions and capturing the program state~\cite{25}.  Alternatively, the Python code is translated to an asynchronous program by injecting \lstinline|async| and \lstinline|await| commands into the program code (cf.\ Baxter et al.~\cite{10}).  Our approach differs significantly from both of these alternatives.

Finally, in some circumstances, the student's program code is not meant to be executed on the computer directly, but rather on another device such as a single-board computer.  MakeCode~\cite{9} and the micro:bit Python Editor~\cite{16,17} therefore do not inherently rely on running Python code in the browser, but download it to, e.g., a micro:bit or Calliope mini using WebUSB for execution.  Nonetheless, the micro:bit editor also supports execution of code inside the browser with a simulator that is based on compiling MicroPython~\cite{19} to WebAssembly.  Since robotics based on micro:bit and Calliope mini is frequently used by schools in our area, \WebTP{} also supports downloading program code to such a device and execute it there---or even transfer the code on the single click of a button if the browser supports WebUSB---, borrowing from the micro:bit Python editor.

\begin{figure*}[t]
\begin{subfigure}{0.32\linewidth}
  \begin{center}
    \begin{tikzpicture}
      \node at (0,0) {\includegraphics[width=.99\linewidth]{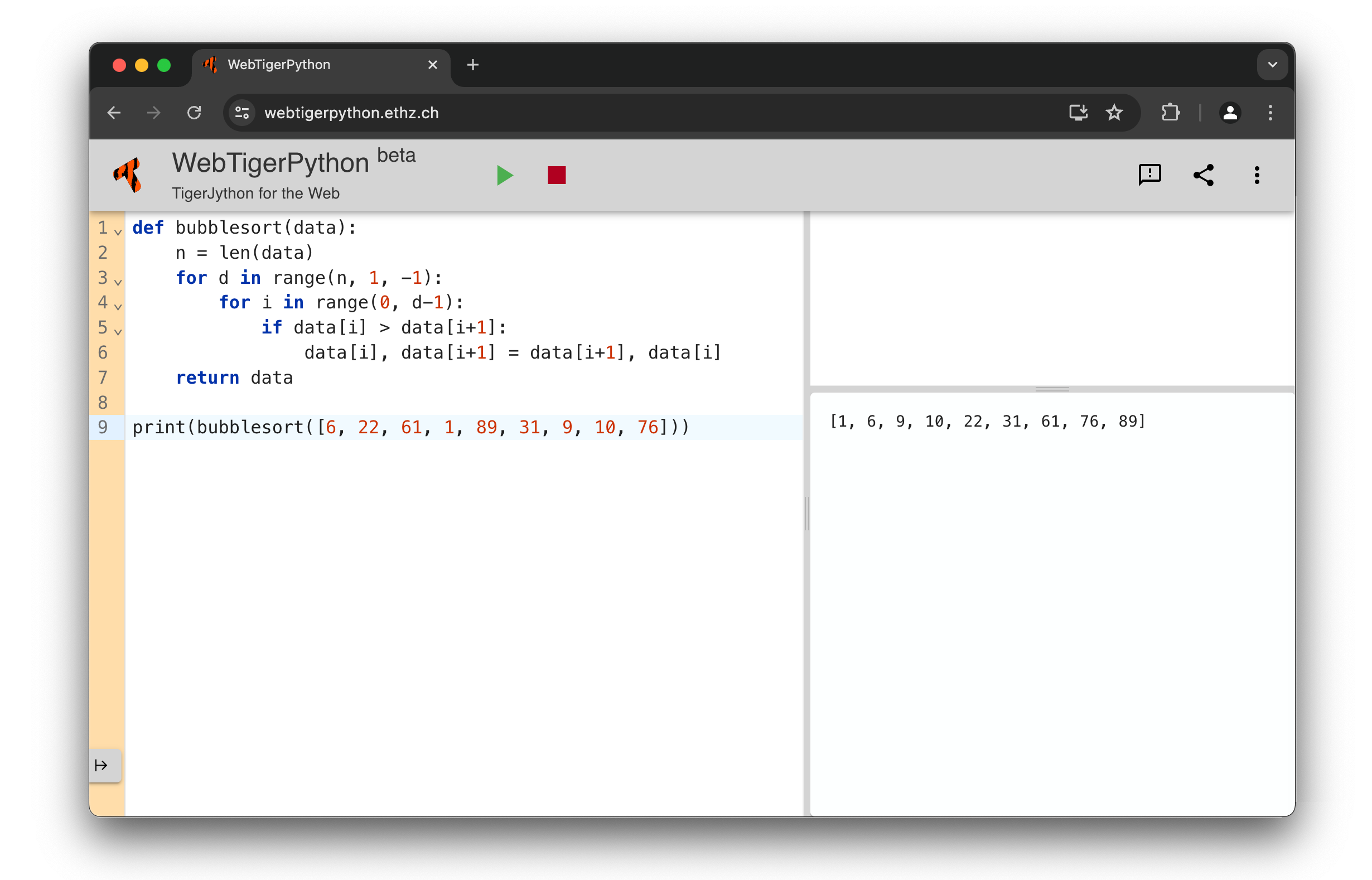}};
    \end{tikzpicture}
    \caption{A simple implementation of Bubblesort}
    \label{fig:screenshot-1}
  \end{center}
\end{subfigure}
\hfill
\begin{subfigure}{0.32\linewidth}
  \begin{center}
    \begin{tikzpicture}
      \node at (0,0) {\includegraphics[width=.99\linewidth]{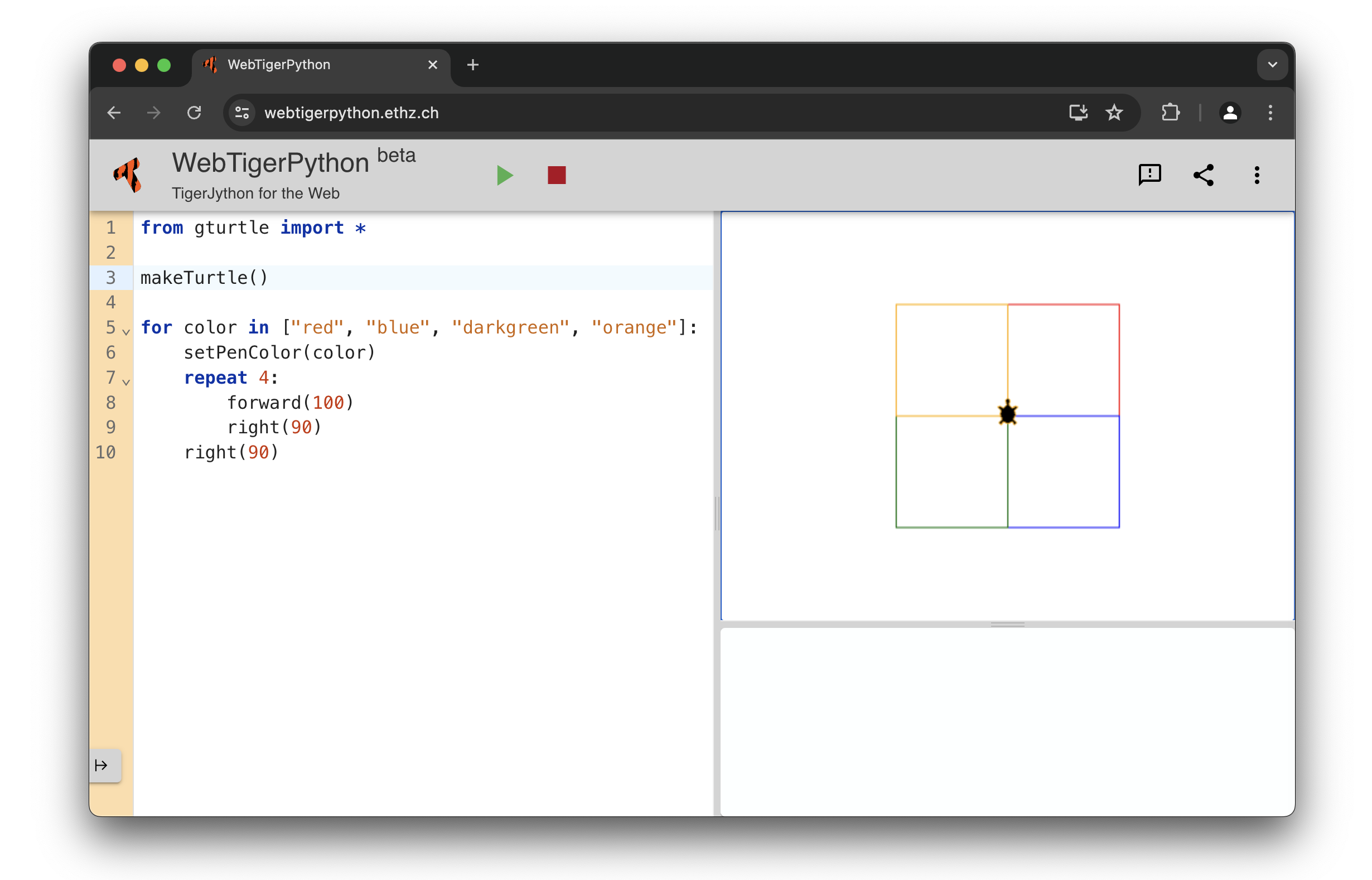}};
    \end{tikzpicture}
    \caption{Drawing shapes with the Turtle}
    \label{fig:screenshot-2}
  \end{center}
\end{subfigure}
\hfill
\begin{subfigure}{0.32\linewidth}
  \begin{center}
    \begin{tikzpicture}
      \node at (0,0) {\includegraphics[width=.99\linewidth]{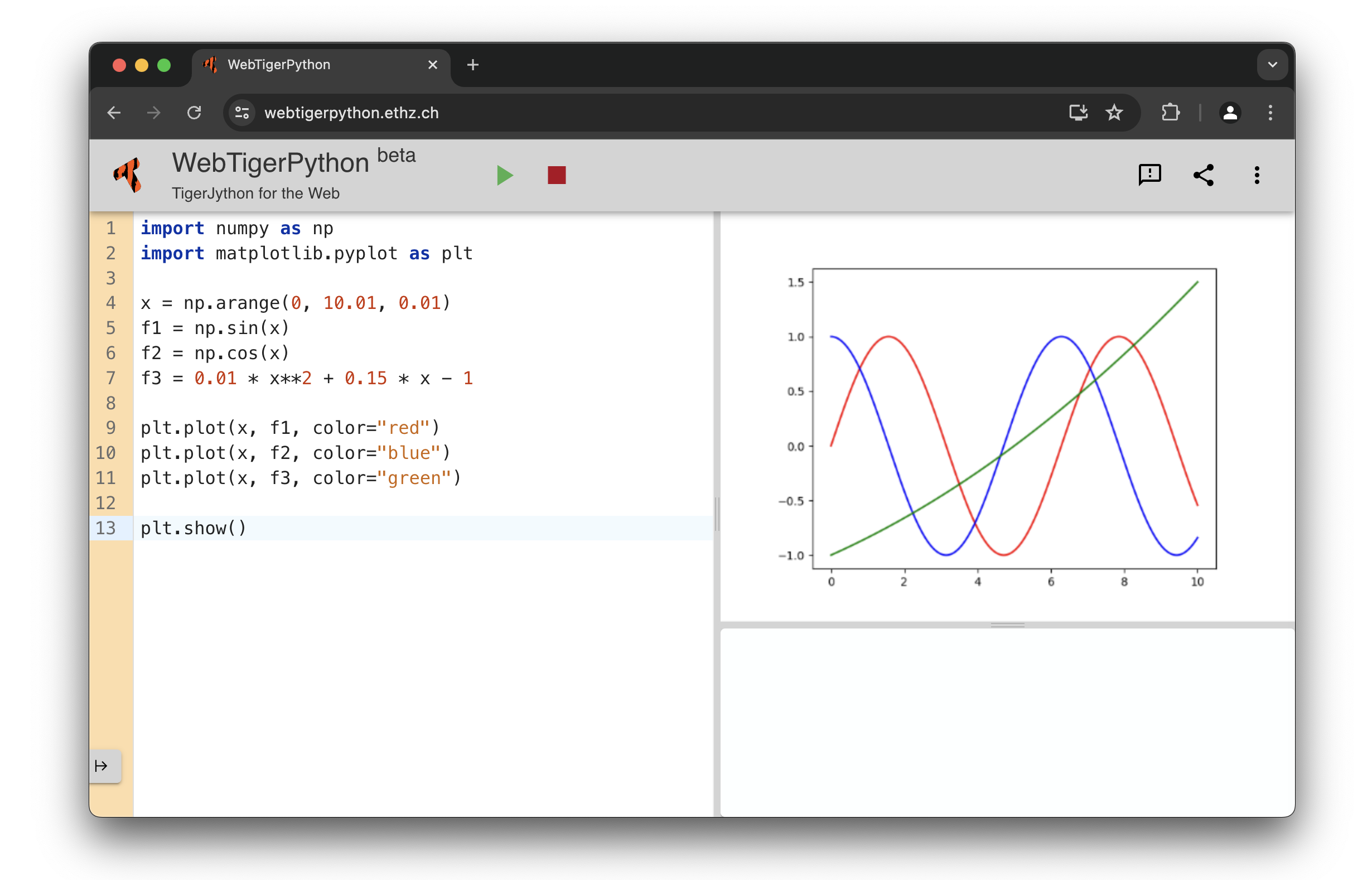}};
    \end{tikzpicture}
    \caption{Plotting functions with Matplotlib}
    \label{fig:screenshot-3}
  \end{center}
\end{subfigure}
\caption{Different screenshots of \WebTP{}}
\label{fig:screenshots}
\end{figure*}

\section{Setting and Features}\label{sec:features}

With mandatory computer science classes in high school (grades 7--12) in Switzerland, there is a need for an accessible, novice-friendly programming environment that can be used to teach programming following established curricula~\cite{8,18,24}.

\subsection{Requirements}\label{subsec:requirements}

Due to a history of introducing Python programming in high school of about twelve years, any new system needs to aim for compatibility with existing teaching materials while also addressing current and upcoming issues.  This brought us to the following catalogue of requirements for \WebTP:

\begin{itemize}[leftmargin=1em]

\item
An uncluttered novice-friendly user interface with language localisation and user-friendly, informative error messages that are easy to understand.

\item
Work seamlessly with existing textbooks and learning materials that are already in use in Switzerland.  Differences between the previously used programming environment and \WebTP{} should be kept to a minimum, allowing for a smooth transition.  This includes support for enhanced Turtle graphics, audio output, and physical computing/robotics (see below).

\item
Run on virtually any device, such as laptops, desktop and tablet computers.

\item
Run out of the box without the necessity to sign up or log in, thereby eliminating technical and privacy issues.

\item
Provide Python's data capabilities in form of NumPy or Matplotlib for the emerging topics on data science in the school curriculum.

\item
Make it easy to share code with others and across devices.

\end{itemize}

\noindent
One of the major points of our tool is to have a low entry barrier for novice programmers, while at the same time not imposing a limit on more advanced users.

\subsection{Progressive Learning Pathways}

\begin{figure}[b]
  \begin{center}
    \begin{tikzpicture}
      \node at (0,0) {\includegraphics[width=.99\linewidth]{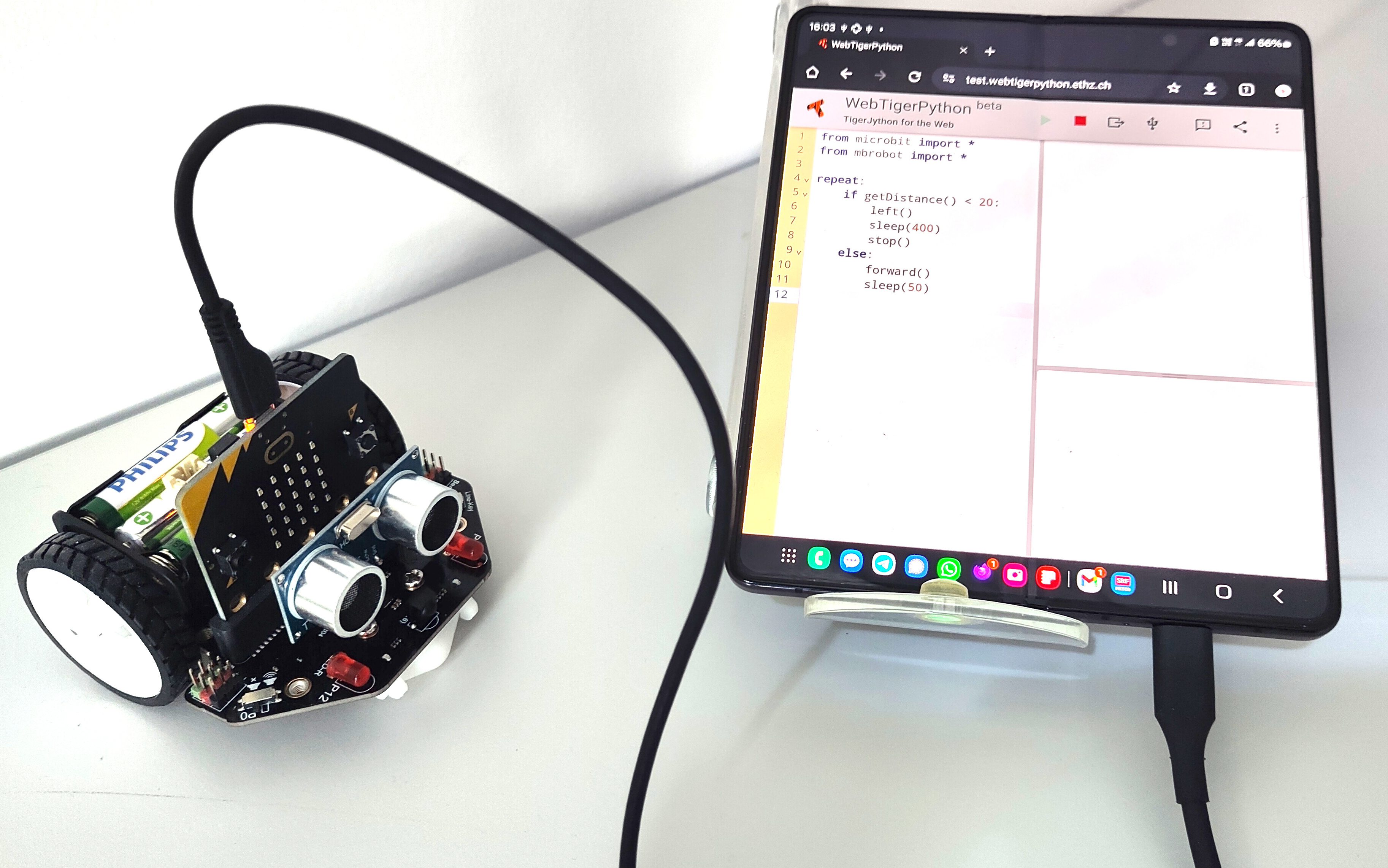}};
    \end{tikzpicture}
  \end{center}
\caption{Programming a Maqueen robot via tablet}
\label{fig:tablet}
\end{figure}

\WebTP{} is designed to support a wide range of educational stages, offering a progressive learning experience that grows with the student's skills, enabling teachers to follow a K--12 spiral curriculum on programming. This is based on features like the following:

\begin{itemize}[leftmargin=1em]

\item \textbf{Built-in Documentation.}
Inspired by block-based IDEs, such as Scratch and the micro:bit Python Editor~\cite{16}, we provide documentation for supported commands in a practical side panel, from where exemplary code snippets can be easily dragged and dropped onto the editor.
    
\item \textbf{Beginner-Friendly Turtle Graphics.} An established way to introduce programming to students revolves around Turtle graphics~\cite{12}.  \WebTP{} offers a tailor-made Turtle library, which enhances Python's standard Turtle (\Cref{SEC:TURTLE-GRAPHICS}). An example is shown in \Cref{fig:screenshot-2}.

\item \textbf{Interactive Graphics.} Keyboard and mouse interactions allow students to deal with input and create more engaging applications, such as, e.g., simple games.

\item \textbf{Physical Computing.} \WebTP{} supports programming with popular single-board computers such as the micro:bit and Calliope mini via WebUSB, along with custom libraries to program the Maqueen and Calli:bot robots, providing hands-on experience with physical computing and educational robotics. \Cref{fig:tablet} shows a Maqueen being programmed from a tablet, for instance.

\item \textbf{Compute-Intensive Applications.} \WebTP{} leverages the compute performance provided by Pyodide, which often is within a factor of $3$--$5\times$ of CPython's performance~\cite{13,48}, and was in the order of around $50\times$ faster than a Skulpt-based implementation.  This is important when dealing with compute-intensive tasks such as, e.g., computing a Mandelbrot set (\Cref{SEC:COMPUTE-BENCHMARKS}).

\item \textbf{Advanced Python Programming.} Being built on Pyodide, \WebTP{} includes a fully fledged Python 3.11 environment supporting many advanced C-based libraries such as SciPy, NumPy, pandas, and Matplotlib---the latter being rendered directly in the browser, as can be seen in \Cref{fig:screenshot-3}. This enables advanced students to tackle more complex topics in data analysis, scientific computing, and data visualisation.

\end{itemize}

\subsection{Enhanced Turtle Graphics}\label{SEC:TURTLE-GRAPHICS}

Turtle graphics is well-known for its animated drawing scheme: students can directly observe how the Turtle is executing the commands and drawing a figure.  However, the teaching materials that serve as a guideline for our implementation go on to disable the animation at some point to then use Turtle graphics for visualising data, creating custom animations, and even simple video games.  As a result, our implementation of Turtle graphics comes with a \emph{direct drawing} mode without Turtle animations that is fast enough for applications such as animations or games.

Furthermore, the greatly varying sizes and pixel densities of screens pose a challenge for Turtle graphics, which usually rests on the assumption that a pixel represents a single step.  Depending on the screen, Turtle images may be inappropriately scaled, not fitting into the available space or being too small.  \WebTP{} therefore features an ``unlimited'' canvas that allows the user to zoom in/out or move the canvas around (using finger gestures on tablet devices).  This is a clear departure from existing systems that have a fixed window, canvas, or image size to work with.

\section{Implementation}\label{sec:tech}

\begin{figure}[t]
  \begin{center}
    \includegraphics[width=\linewidth]{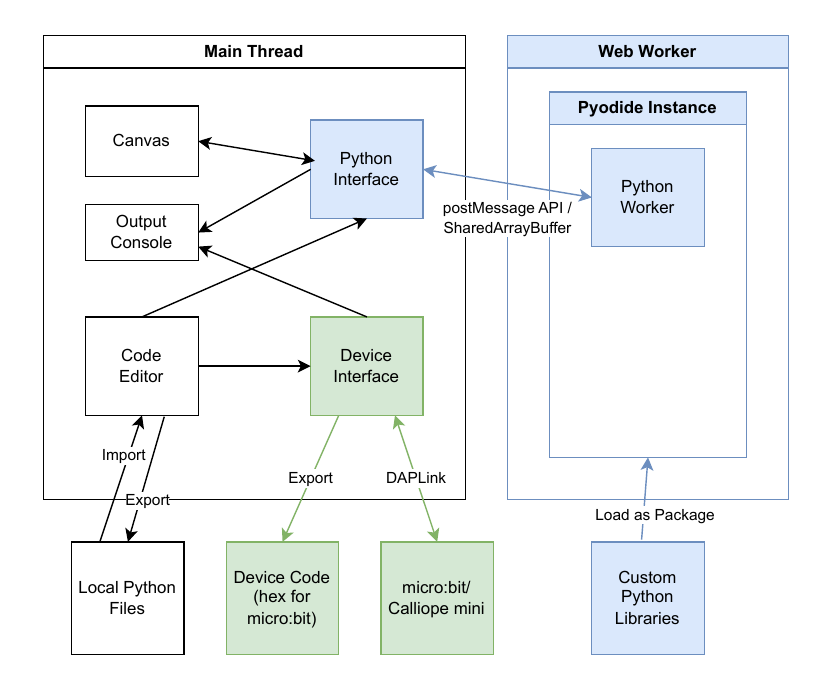}
  \end{center}
  \caption{System overview of \WebTP}
  \label{fig:technology}
\end{figure}

We designed a new custom IDE, which is visually inspired by its predecessor WebTigerJython~\cite{45}. The standard layout includes a code editor, a canvas for rendering graphical output, and an output field; see \Cref{fig:screenshots}. We opted for a simple text output instead of a regular console since input is implemented through an interactive dialogue. When designing the user interface we adhered to common design principles while also trying to make the IDE as touch friendly as possible. It can be used on any device that can run a modern web browser, though we recommend using it with a physical keyboard for the best experience.

In order to facilitate sharing of programs, the program code can be directly encoded in the URL.  A `share' button in the user interface generates the corresponding URL, which can then be sent to other people or devices.  Although this approach is limited to smaller programs, it works without incurring the cost of a database backend on the server.

\subsection{Python Execution}

\WebTP{} works with a \emph{main thread} and a \emph{web worker}; see \Cref{fig:technology}. The main thread manages the user interface, while a web worker handles Python execution in Pyodide. Communication between the web worker and the main thread occurs via the \emph{postMessage} API. As mentioned above, running Pyodide in a web worker is essential since Python execution in WebAssembly is blocking, and thus executing Python in the main thread would freeze the interface during execution, which would harm the user experience (cf.\ Jefferson et al.~\cite{25}).

To interrupt or halt Python execution in the blocked web worker, we use a \emph{shared array buffer}~\cite{32}, a raw binary buffer accessible by both threads. For a specific value in the buffer, the execution pauses. This is necessary for functions such as rendering in the main thread or waiting for user input.

\subsection{Turtle Library}

The original Python Turtle is based on Tkinter, a Python binding of the \emph{Tk GUI toolkit}, which is not usable in the browser. We decided to re-implement the Turtle library from scratch using the web-based graphics library PixiJS~\cite{34}, a JavaScript library for rendering 2D graphics. It is not straightforward to access PixiJS directly from the web worker, adding to the challenge of communication between the web worker and the main thread.

We implemented most of the Turtle functionality directly in PixiJS. Since PixiJS is a very fast and highly optimised JavaScript rendering library, we decided to create an API that can be called from Python within our web worker. To minimise communication overhead between the two threads, multiple instructions can be bundled together. This bundling is crucial if the Turtle is used to render custom animations or games that require fast drawing. Due to the communication overhead between the threads, the animations would be lagging without bundling.

\subsection{Interactivity}

Users can interact with their programs either through the \lstinline|input()| function, or via mouse and keyboard events. The former opens a GUI dialogue, prompting the user to enter a string which is then sent to Python and can be used within the program. Instead of the common browser prompt dialogue, we use a custom input dialogue which does not block the execution within the main thread like the regular input function does.

Additionally, we log keyboard events and interactions with the canvas. Events can be queried from Python in both a blocking and a non-blocking way. It is also possible to register callback functions that listen for events after program execution. However, a limitation is that callback functions cannot be executed during Python execution as Python is single-threaded. If events are triggered during the execution of Python code, they are registered in an event queue and processed one by one as soon as the execution ends.

\subsection{Data and Function Plotting}

Although Matplotlib comes with an HTML 5 backend~\cite{43}, we were not able to use it as we run Python in a web worker, which has no access to the DOM. To still be able to render plots, we use the AGG (Anti Grain Geometry) backend~\cite{1}, a general purpose graphics toolkit available in Pyodide.

To display the image on the canvas, we first render it to a buffer within Python and then pass it to the main thread, where we display it on our canvas; a result is shown in \Cref{fig:screenshot-3}. This allows Matplotlib and other libraries that build on it, such as Seaborn and NetworkX, to be used in \WebTP{}. However, animations and interactive features of Matplotlib are not yet supported.

\subsection{Physical Computing}

Single-board computers such as the micro:bit or Calliope mini can also be programmed with \WebTP{}. The Python code can be exported with MicroPython and loaded onto the external computer. Using WebUSB, it is even possible to transfer the program directly using the browser on a single click.

The external computer can communicate with the IDE over a serial connection with the main thread, which is similar to the code execution in the web worker. This can then be used to display error messages and other code output that cannot be shown properly on the external computer.

There are also extension boards for robots such as the Maqueen or the Calli:bot, which can be programmed via pins. This process is cumbersome and unsuitable for novices. Therefore, we provide additional abstraction libraries~\cite{31}, which are also flashed to the single-board computer if needed.

\section{Evaluation}\label{sec:eval}

We were able to implement all of the functionality necessary in order to follow the established programming curriculum.  Currently, \WebTP{} has an average number of around 500 unique users per day (and up to roughly 1\,000), where ``unique'' is defined as having the same IP for 24 hours.

\subsection{Compute Benchmarks}\label{SEC:COMPUTE-BENCHMARKS}

In order to test the potential for compute-intensive applications, we computed a simple Mandelbrot set for an image of size $200\times 200$ pixels, leading to the computation
$(z' = z^2\ \text{+}\ c$
being executed 2\,000\,000 times.  In order to focus on the compute performance, we did not display the resulting image.  This benchmark was conducted on an AMD Threadripper 7960X.

CPython needed $0.14\,\mathrm{s}$, \WebTP{} $0.38\,\mathrm{s}$, PyodideU $107.22\,\mathrm{s}$, and Trinket.io $16.88\,\mathrm{s}$ (best of three runs, variance in all cases within $1\,\%$).  \WebTP{} comes within $3\times$ of CPython's performance and is about $44\times$ faster than Skulpt-based Trinket.io.  Particularly noteworthy is PyodideU's rather slow performance, which is in spite of also using Pyodide as the Python interpreter.  In contrast to \WebTP, PyodideU runs the code inside the main thread.  In order to keep the UI responsive, execution is regularly interrupted and resumed using exceptions that capture the current state~\cite{25}.  According to our simple benchmark this incurs a massive performance penalty of nearly $300\times$.

\subsection{Graphics Benchmarks}\label{subsection:benchmarks}

To check the performance of the Turtle library we wanted to ensure that animations still work smoothly in our IDE. The Pyodide team states that Pyodide is $3$--$5\times$ slower than running Python locally~\cite{13,48}; however, since most of the rendering happens in PixiJS, we were confident that it would be possible to get a similar rendering performance.

\begin{figure}[!t]
  \begin{subfigure}{0.9\linewidth}
    \begin{tikzpicture}
\begin{axis}[
    width=\textwidth,
    height=0.6\textwidth,
    xlabel={number of balls},
    ylabel={frames per second},
    xmode=log,
    ymode=log,
    log basis x=10,
    log basis y=10,
    label style={font=\small},
    legend pos={south west},
    legend style={font=\scriptsize,cells={anchor=west}},
    grid=both,
    minor grid style={gray!25},
    major grid style={gray!50},
    xtick={10, 20, 50, 100, 200, 500, 1000},
    xticklabels={10, 20, 50, 100, 200, 500, 1000},
    ymin=0.1, ymax=1000,
    cycle list name=color list
]

\addplot+[black,mark=*] table {
    x y
    10 410
    20 233
    50 114
    100 60
    200 31
    500 12.4
    1000 5.7
};
\addlegendentry{Python + Tkinter}

\addplot+[blue!70,mark=square*] table {
    x y
    10 374
    20 215
    50 151
    100 90
    200 45
    500 18
    1000 9.1
};
\addlegendentry{\WebTP{}}

\addplot+[red!70,mark=triangle*] table {
    x y
    10 60
    20 60
    50 60
    100 60
    200 36
    500 5.7
    1000 0.91
};
\addlegendentry{Skulpt}

\end{axis}
\end{tikzpicture}
    \caption{Benchmark 1 using bouncing balls}
    \label{fig:benchmark-1}
  \end{subfigure}\\[4mm]
  \begin{subfigure}{0.9\linewidth}
    \begin{tikzpicture}
\begin{axis}[
    width=\textwidth,
    height=0.6\textwidth,
    xlabel={number of birds},
    ylabel={frames per second},
    xmode=log,
    ymode=log,
    log basis x=10,
    log basis y=10,
    label style={font=\small},
    legend pos={south west},
    legend style={font=\scriptsize,cells={anchor=west}},
    grid=both,
    minor grid style={gray!25},
    major grid style={gray!50},
    xtick={10, 20, 50, 100, 200, 500, 1000},
    xticklabels={10, 20, 50, 100, 200, 500, 1000},
    ymin=0.1, ymax=1000,
    cycle list name=color list
]

\addplot+[black,mark=*] table {
    x y
    10 117
    20 61
    50 25
    100 13
    200 6
    500 2.4
    1000 1.1
};
\addlegendentry{Python + Tkinter}

\addplot+[blue!70,mark=square*] table {
    x y
    10 300
    20 134
    50 48
    100 21
    200 9
    500 5
    1000 1.3
};
\addlegendentry{\WebTP{}}

\addplot+[red!70,mark=triangle*] table {
    x y
    10 60
    20 51
    50 24
    100 12
    200 5.9
    500 2.4
    1000 0.7
};
\addlegendentry{Skulpt}

\end{axis}
\end{tikzpicture}
    \caption{Benchmark 2 using a simple animated bird}
    \label{fig:benchmark-2}
  \end{subfigure}
  \caption{Benchmarks with animated objects}
  \label{fig:benchmarks}
\end{figure}

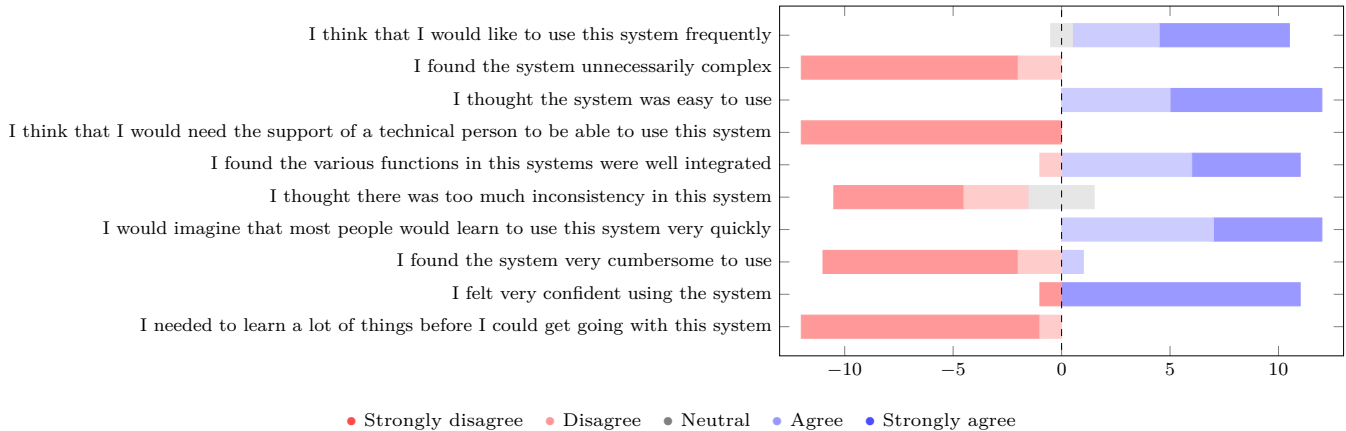
\begin{figure*}[!t]
\pgfplotstableread{
Question                                                                                      SD DI NE AG SA
{I needed to learn a lot of things before I could get going with this system}                 11  1  0  0  0
{I felt very confident using the system}                                                       1  0  0  0  11
{I found the system very cumbersome to use}                                                    9  2  0  1  0
{I would imagine that most people would learn to use this system very quickly}                 0  0  0  7  5
{I thought there was too much inconsistency in this system}                                    6  3  3  0  0
{I found the various functions in this systems were well integrated}                           0  1  0  6  5
{I think that I would need the support of a technical person to be able to use this system}   12  0  0  0  0
{I thought the system was easy to use}                                                         0  0  0  5  7
{I found the system unnecessarily complex}                                                    10  2  0  0  0
{I think that I would like to use this system frequently}                                      0  0  1  4  6
}\testdata
\begin{center}
  \begin{tikzpicture}[scale=0.875]
    \begin{axis}[
      height=6.9cm,
      width=10.15cm,
      legend cell align=left,
      legend columns=6,
      legend style={at={(0.5,-0.15)},anchor=north,draw=none},
      xbar stacked,
      stack negative=separate,
      xmin=-13,
      xmax=13,
      xtick distance=5,
      xticklabel=,
      ytick=data,
      yticklabels from table={\testdata}{Question},
      ticklabel style = {font=\small}
      ]
      \addlegendimage{empty legend}
      \draw[dashed] (axis cs: 0,-5)--(axis cs: 0,15);
      \addplot[black!10, fill=black!10] table [x expr=\thisrow{NE}/-2, meta=Question, y expr=\coordindex] {\testdata};
      \addplot[red!20, fill=red!20] table [x expr=-\thisrow{DI}, meta=Question, y expr=\coordindex] {\testdata};
      \addplot[red!40, fill=red!40] table [x expr=-\thisrow{SD}, meta=Question, y expr=\coordindex] {\testdata};
      \addplot[black!10, fill=black!10] table [x expr=\thisrow{NE}/2, meta=Question, y expr=\coordindex] {\testdata};
      \addplot[blue!20, fill=blue!20] table [x expr=\thisrow{AG}, meta=Question, y expr=\coordindex] {\testdata};
      \addplot[blue!40, fill=blue!40] table [x expr=\thisrow{SA}, meta=Question, y expr=\coordindex] {\testdata};
    \end{axis}
    \node[font=\footnotesize] (1) at (-1.5,-1) {\textcolor{red!70}{$\bullet$} Strongly disagree\quad\textcolor{red!40}{$\bullet$} Disagree\quad\textcolor{black!50}{$\bullet$} Neutral\quad\textcolor{blue!40}{$\bullet$} Agree\quad\textcolor{blue!70}{$\bullet$} Strongly agree};
  \end{tikzpicture}
  \caption{Accumulated answers of each item of the \textit{System Usability Scale} questionnaire}
  \label{fig:sus}
\end{center}
\end{figure*}

We compared \WebTP{} to two other Turtle implementations: the reference CPython Turtle implementation with a Tkinter backend and a Skulpt-based Turtle running in the Trinket.io IDE~\cite{6}, which is another web-based implementation of the Turtle using an HTML canvas element. A major difference between our implementation and the Skulpt Turtle is that the latter runs in the main thread and therefore avoids communication overhead.

As a benchmark, we ran two animations, one with bouncing balls, and one with minimalistic ``birds'' flapping their wings. The benchmarks were all run within an area of $900\times 900$ pixels, conducted on an Intel Core i7-8750H. We ran the exact same code on all Python implementations.  The results are shown in \Cref{fig:benchmarks}, demonstrating that we were able to achieve better performance than both the Python and the Skulpt implementations (note, however, that the Skulpt Turtle is capped at 60~fps and therefore not able to achieve a higher frame rate).

\subsection{Turtle Comparison}\label{subsection:turtle_rendering}

While all Turtle implementations produce similar graphics, there are a few differences to note. The standard Python Turtle with a Tkinter backend does not run in the browser. The Skulpt Turtle runs in browsers and uses an HTML canvas as a backend. Although it covers all features of the Python Turtle, it runs in Skulpt---an implementation of Python in JavaScript—lacking some Python functionality. The Brython Turtle uses an SVG backend. However, many Turtle functions, including animation control, are missing, making it incomparable to our benchmark.
An overview of features of different Turtle implementations is shown in \Cref{tab:turtles}.

The feature of zooming and panning is part of our Turtle and not part of any of the other Turtle implementations.

\begin{table}[h]
\renewcommand{\arraystretch}{1.1}\setlength\tabcolsep{0.25em}
\small
\caption{Comparison of different Turtle implementations}
\label{tab:turtles}
\begin{tabular}{@{}r c c c c@{}}
  \toprule
  Turtle in & Python & Brython & Skulpt & \WebTP{} \\
  \midrule
  Web &  & \checkmark & \checkmark & \checkmark \\ 
  Runs in CPython & \checkmark & & & \checkmark \\
  Interactive Execution & \checkmark & & \checkmark & \checkmark \\
  Events & \checkmark & & \checkmark &\checkmark \\
 \bottomrule
\end{tabular}
\end{table}

\subsection{Usability Survey}

In order to get some first feedback, we conducted a small survey.  To this end, we used our network of in-service teachers to ask regular users of \WebTP{} to fill out the \emph{System Usability Scale}~\cite{11}.  An average score of 89 (of at most 100) was achieved, with 12 teachers taking part in the survey, which can be interpreted as a very good outcome~\cite{40}; the results are shown in \Cref{fig:sus}.

\section{Limitations and Future Work}\label{sec:limits}

Within the diverse and fast evolving landscape of web applications, your experience may vary considerably, both in terms of performance and features.  For instance, WebUSB is currently only supported by Chrome and Edge, and thus not available on iOS, say.  Moreover, since we are executing the Python programs inside an isolated web worker without access to the user interface, packages like Pygame~\cite{35} are not compatible, although they are supported by Pyodide.  The solution might be a hybrid similar to PyodideU.  However, with discussions about making OpenGL accessible from within web workers and extensions to WebAssembly, the possibilities of web workers might expand significantly in the coming months and years.

Work is also ongoing to improve basic functionality, implement further Python modules, and add support for integration with classroom management systems.  While support for multiple files is possible in principle, it is not yet exposed in the current user interface.  However, the current implementation is mature enough to be used in classroom and serves as a basis for research in computing education and usability.

The current evaluation demonstrates that existing textbooks in Switzerland can be used with \WebTP{}, and we have received some encouraging feedback from teachers.  Nonetheless, a full evaluation of the system with extensive classroom experience will be conducted over the coming months.

\section{Conclusion}\label{sec:conclusion}

With our prototype of \WebTP{}, we were able to fulfil all requirements formulated in \Cref{subsec:requirements}.  We are therefore positive that it will find wide adaptation by teachers in Switzerland and beyond.  In particular, \WebTP{} allows for a low floor (e.g., Turtle graphics, educational robotics) and a high ceiling (in particular advanced CPython libraries).

For our purposes, running Pyodide in a web worker turned out to be surprisingly efficient as is demonstrated by a set of preliminary benchmarks.

\section*{Acknowledgements}

The authors thank C\'edric Donner for help with the design choices, Julia Bogdan for the initial Turtle prototype, and Daniel Barot for the canvas interactivity. More on the history of \WebTP{} can be found in a recent blog post~\cite{30}.

\end{document}